\documentclass[draft,jgrga]{agutex}

\usepackage{lineno}
\linenumbers*[1]

  \usepackage[dvips]{graphicx}
  \setkeys{Gin}{draft=false}

\authorrunninghead{KRIVOVA ET AL.}

\titlerunninghead{SOLAR UV IRRADIANCE SINCE 1974}

\authoraddr{N. A. Krivova,
Max-Planck-Institut f\"ur Sonnensystemforschung,
Max-Planck-Str. 2, 37191 Katlenburg-Lindau, Germany
(natalie@mps.mpg.de)}

\authoraddr{S. K. Solanki,
Max-Planck-Institut f\"ur Sonnensystemforschung,
Max-Planck-Str. 2, 37191 Katlenburg-Lindau, Germany}

\authoraddr{T. Wenzler,
Hochschule f\"ur Technik Z\"urich, CH-8004 Z\"urich,
Switzerland}

\authoraddr{B. Podlipnik,
Max-Planck-Institut f\"ur Sonnensystemforschung,
Max-Planck-Str. 2, 37191 Katlenburg-Lindau, Germany}

\begin{document}

\title{Reconstruction of solar UV irradiance since 1974}

\authors{N. A. Krivova, \altaffilmark{1}
S. K. Solanki, \altaffilmark{1,}\altaffilmark{2}
T. Wenzler, \altaffilmark{1,}\altaffilmark{3}
and B. Podlipnik \altaffilmark{1}}

\altaffiltext{1}{Max-Planck-Institut f\"ur Sonnensystemforschung,
D-37191 Katlenburg-Lindau, Germany}

\altaffiltext{2}{School of Space Research, Kyung Hee University,
Yongin, Gyeonggi 446-701, Korea}

\altaffiltext{3}{Hochschule f\"ur Technik Z\"urich, CH-8004 Z\"urich,
Switzerland}

\begin{abstract}

Variations of the solar UV irradiance are an important driver of chemical
and physical processes in the Earth's upper atmosphere and may also
influence global climate.
Here we reconstruct solar UV irradiance in the range 115--400~nm over the
period 1974--2007 by making use of the recently developed empirical
extension of the SATIRE models employing SUSIM data.
The evolution of the solar photospheric magnetic flux, which is a central
input to the model, is described by the magnetograms and continuum images
recorded at the Kitt Peak National Solar Observatory between 1974 and 2003
and by the MDI instrument on SoHO since 1996.
The reconstruction extends the available observational record by 1.5~solar
cycles.
The reconstructed Ly-$\alpha$ irradiance agrees well with the composite time
series by \citet{woods-et-al-2000a}.
The amplitude of the irradiance variations grows with decreasing wavelength
and in the wavelength regions of special interest for studies of the Earth's
climate (Ly-$\alpha$ and oxygen absorption continuum and bands between 130
and 350~nm) is one to two orders of magnitude stronger than in the visible
or if integrated over all wavelengths (total solar irradiance).
\end{abstract}

\begin{article}

% main text
\section{Introduction}
\label{intro}

Solar irradiance variations show a strong wavelength dependence.
Whereas the total (integrated over all wavelengths) solar irradiance (TSI)
changes by about 0.1\% over the course of the solar cycle, the irradiance in
the UV part of the solar spectrum varies by up to 10\% in the 150--300~nm
range and by more than 50\% at shorter wavelengths, including the
Ly-$\alpha$ emission line near 121.6~nm \citep[e.g.][]{floyd-et-al-2003a}.
On the whole, more than 60\% of the TSI variations over the solar cycle are
produced at wavelengths below 400~nm [\citealp{krivova-et-al-2006a}; cf.
\citealp{harder-et-al-2009}].

These variations may have a significant impact on the Earth's climate
system.
Ly-$\alpha$, the strongest line in the solar UV spectrum, which is formed
in the transition region and the chromosphere, takes an active part in
governing the chemistry of the Earth's upper stratosphere and mesosphere,
e.g., by ionizing nitric oxide, which affects the electron density
distribution, or by stimulating dissociation of water vapor and producing
chemically active HO(x) that destroy ozone
\citep[e.g.][]{frederick-77,brasseur-simon-81,huang-brasseur-93,%
fleming-et-al-95,egorova-et-al-2004,langematz-et-al-2005a}.
Also, radiation in the Herzberg oxygen continuum (200--240~nm) and the
Schumann-Runge bands of oxygen (180--200~nm) is important for photochemical
ozone production \citep[e.g.][]{haigh-94,haigh-2007,egorova-et-al-2004,%
langematz-et-al-2005b,rozanov-et-al-2006,austin-et-al-2008}.
UV radiation in the wavelength range 200--350~nm, i.e. around the
Herzberg oxygen continuum and the Hartley-Huggins ozone bands, is the
main heat source in the stratosphere and mesosphere
\citep{haigh-99,haigh-2007,rozanov-et-al-2004,rozanov-et-al-2006}.

The record of regular measurements of the solar UV irradiance spectrum,
accurate enough to assess its variations, goes back to 1991, when the Upper
Atmosphere Research Satellite (UARS) was launched.
Among others, it carried two instruments for monitoring solar radiation in
the UV, the Solar Ultraviolet Spectral Irradiance Monitor
\cite[SUSIM;][]{brueckner-et-al-93} and the Solar Stellar Irradiance
Comparison Experiment \citep[SOLSTICE;][]{rottman-et-al-93}.
These data sets are of inestimable value, but remain too short to allow
reliable evaluation of solar influence on the Earth's climate and need to be
extended back in time with the help of models.

Reconstructions of solar UV irradiance have earlier been presented by
\cite{fligge-solanki-2000} and by \cite{lean-2000}.
The first one was based on LTE (Local Thermodynamic Equilibrium)
calculations of the solar spectrum and the latter on UARS/SOLSTICE
measurements.
The LTE approximation gives inaccurate results below approximately 200~nm
and in some spectral lines, whereas the long-term uncertainty of SOLSTICE
(as well as of all other instruments that measured solar UV irradiance
before SORCE) exceeded the solar cycle variation above approximately 250~nm,
thus leading to incorrect estimates of the UV irradiance variability at
longer wavelengths \citep[see][]{lean-et-al-2005,krivova-et-al-2006a}. 

Whereas considerable advance has recently
been made in modelling the variations of
the total solar irradiance and the irradiance at wavelengths longer than
about 300~nm \citep[e.g.,][]{unruh-et-al-99,ermolli-et-al-2003,%
krivova-et-al-2003a,wenzler-et-al-2004a,wenzler-et-al-2005a,wenzler-et-al-2006a},
models at shorter wavelengths have not kept apace.
This is because the LTE approximation usually taken in calculations of the
brightness of different photospheric components fails in this wavelength
range and non-LTE calculations are much more arduous
\citep[e.g.][]{fontenla-et-al-99,fontenla-et-al-2006,%
haberreiter-et-al-2005}.

An alternative approach has been developed by
\citet{krivova-solanki-2005a} and \citet {krivova-et-al-2006a} that allows
an empirical extrapolation of the successfull SATIRE models
\citep{krivova-solanki-2005b,solanki-et-al-2005a} down to 115~nm using
available SUSIM measurements.
\citet{krivova-et-al-2006a} have combined this technique with the model of
\citet{krivova-et-al-2003a} to reconstruct the variations of the solar UV
irradiance over the period 1996--2002, i.e. the rising phase of cycle~23,
using MDI (Michelson Doppler Imager on SoHO) \citep{scherrer-et-al-95}
magnetograms and continuum images.
Here we employ the data from the National Solar Observatory Kitt Peak (NSO KP),
in order to reconstruct the solar UV irradiance spectrum back to 1974.
We then combine this KP-based reconstruction for the period 1974--2002 with
the reconstruction based on MDI data \citep{krivova-et-al-2006a}, which has
now been extended to 2006.
In order to fill in the gaps in daily data and to extend the time series to
2007, when MDI continuum images displayed deteriorating quality, we employ
the Mg~II core-to-wing ratio and the solar F10.7~cm radio flux.
Hence the present paper extends the work of \citet{krivova-et-al-2006a} to
three cycles, i.e. the whole period of time over which high quality
magnetograms are available.

The model is described in Sect.~\ref{model}, the results are presented
in Sect.~\ref{results} and summarised in Sect.~\ref{summary}.

\section{Model}
\label{model}

We take a similar approach as \citet{krivova-et-al-2006a}.
This means that variations of the solar total and spectral irradiance on
time scales of days to decades are assumed to be entirely due to the
evolution of the solar surface magnetic field.
Under this assumption, \citet{krivova-et-al-2003a} and
\citet{wenzler-et-al-2005a,wenzler-et-al-2006a} have 
successfully modelled
the observed variations of the total solar irradiance.
\citet{krivova-et-al-2006a} showed that this (SATIRE) model also works well
in the spectral range 220--240~nm (hereinafter, the reference range).
They then analysed SUSIM data and worked out empirical relationships between
the irradiance in this range and irradiances at all other wavelengths
covered by the SUSIM detectors (115--410~nm).
Thus if the irradiance in the range 220--240~nm is known, it is also
possible to calculate irradiance at other wavelengths in the UV down to 115~nm.

\subsection{Solar irradiance at 220--240~nm}

In a first step, we apply the SATIRE model
\citep[Spectral and Total Irradiance
REconstructions,][]{solanki-et-al-2005a,krivova-solanki-2008a} to NSO
KP magnetograms and continuum images, in order to reconstruct solar
irradiance in the reference range for the period 1974--2003.
In SATIRE,
the solar photosphere is divided into 4 components: the quiet Sun, sunspot
umbrae, sunspot penumbrae and bright magnetic features (describing both
faculae and the network).
Each component is decribed by the time-independent spectrum calculated from
the corresponding model atmospheres in the LTE approximation
\citep{unruh-et-al-99}.
Since the distribution of the magnetic field on the solar surface evolves
continuously, the area covered by each of the components on the visible
solar disc also changes.
This is represented by the corresponding filling factors, which are
retrieved from the magnetograms and continuum images.
In the period 1974--2003, such data were recorded (nearly daily) with the
512-channel Diode Array Magnetograph \citep[before
1992;][]{livingston-et-al-76} and the Spectromagnetograph
\citep[after 1992;][]{jones-et-al-92} on Kitt Peak
\citep[see also][]{wenzler-et-al-2006a}.
Since 1996 magnetograms and continuum images were also recorded by the MDI
instrument on SoHO.
More details about the SATIRE model have been given by
\citet{fligge-et-al-2000a,krivova-et-al-2003a,wenzler-et-al-2005a,%
wenzler-et-al-2006a}.

This model has one free parameter, $B_{\rm sat}$, denoting the field
strength below which the facular contrast is proportional to the magnetogram
signal, while it is independent (saturated) above that.
It depends on the quality (noise level and spatial resolution) of the
employed magnetograms.
From a comparison with the PMOD composite \citep{froehlich-2006} of the TSI
measurements, \citet{wenzler-et-al-2006a,wenzler-et-al-2009a} found the
value of $B_{\rm sat}=320$~G for the KP data, whereas
\citet{krivova-et-al-2003a} obtained a value of $B_{\rm sat}=280$~G for the
MDI data.
In this work we use the same values of this parameter and do not vary them
any more in order to fit the spectral data.

\begin{figure*}
 \noindent
\includegraphics[width=39pc]{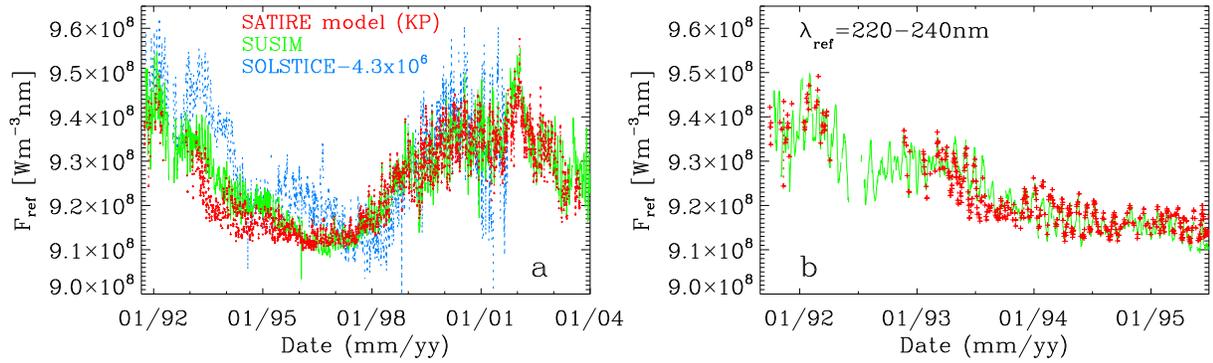}
 \caption[]{\label{refran}
{\bf (a)} The solar irradiance integrated over the wavelength range
220--240~nm as a function of time for the period 1991--2003.
The green line shows SUSIM measurements \citep{floyd-et-al-2003b} and
the red plus signs (connected by the dashed line where there are no gaps)
the values reconstructed using SATIRE models and KP magnetograms.
SOLSTICE data \citep{woods-et-al-96} shifted by $-4.3\times
10^6$\,Wm$^{-3}$nm are shown by the blue dashed line.
{\bf (b)} Enlargement of the panel a restricted to the period before 1996
showing only SUSIM data and the reconstruction.
Here SUSIM data were shifted by $-5.0\times 10^6$\,Wm$^{-3}$nm.
}
\end{figure*}
\begin{figure}
 \noindent
\includegraphics[width=20pc]{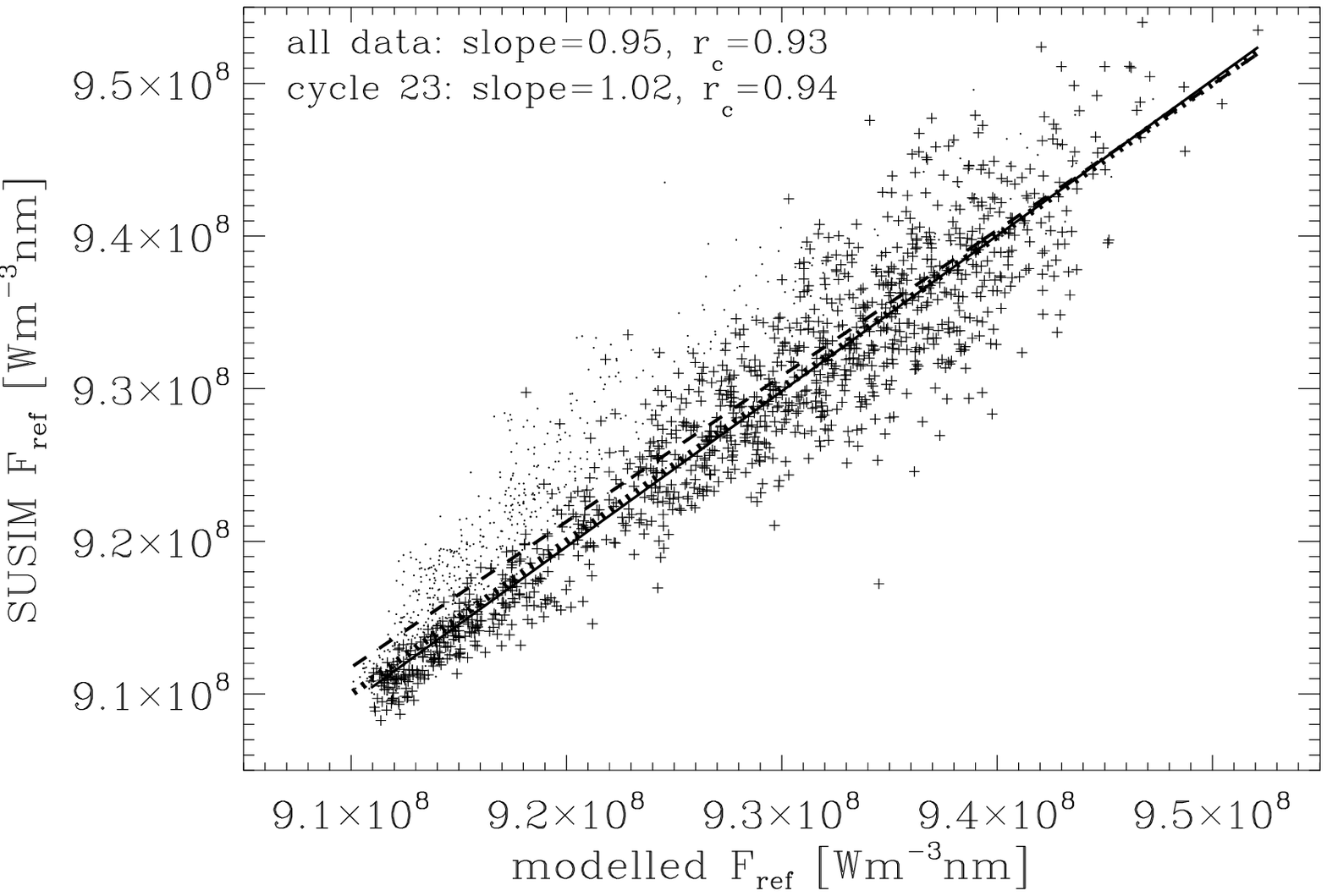}
 \caption[]{\label{fit}
The solar irradiance in the range 220--240~nm: as measured by SUSIM vs.
reconstructed by SATIRE.
Dots and pluses are used for cycles~22 and 23, respectively.
The dashed straight line is the regression to all points (with no correction
applied to SUSIM's absolute level).
The solid line is the regression for cycle~23 only. 
The thick dotted line almost coinciding with the solid line shows the
expectation value, i.e. a slope of 1.0 and no offset.
Correlation coefficients and slopes are indicated in the figure.
}
\end{figure}

The solar irradiance integrated over the wavelength range
220--240~nm reconstructed from the KP magnetograms and continuum images
is shown in Fig.~\ref{refran} by the red plus signs connected by the
dashed line where there are no gaps in the daily sequence of data.
The measurements by the SUSIM instrument are represented by the green line.
We use daily level~3BS V22 data with a sampling of 1~nm
(\citealp{floyd-et-al-2003b}; L.~Floyd, priv. comm.).
A similar plot obtained with MDI data was given by
\citet{krivova-et-al-2006a}.
The apparent change in the behavior between cycles~22 and 23 seen in
Fig.~\ref{refran}a is due to the incorrect estimate of the degradation
during the solar minimum period (L.~Floyd, priv. comm.).
This is, for example, confirmed by a comparison with the Mg~II
core-to-wing ratio, which is free of such problems, and is discussed in more
detail by \citet{krivova-et-al-2006a}.
The fact that a single shift in absolute values applied to the SUSIM data
before 1996 is sufficient in order to bring the data in agreement with the
model also supports this conjecture.
Indeed, in Fig.~\ref{refran}b the period before 1996 is shown on an enlarged
scale.
Here the measurements by SUSIM were shifted in the absolute level by a fixed
value ($-5.0\times 10^6$\,Wm$^{-3}$nm), and a good correspondence between the model
and the data is seen.

This is also demonstrated by Fig.~\ref{fit}, where the measured irradiance
at 220--240~nm is plotted against the modelled values.
Dots and plus signs are used for the data from cycles~22 and 23, respectively
(no correction to the absolute level has been applied).
The dashed straight line with a slope of 0.95 represents the regression to
all points.
The correlation coefficient is 0.93.
The solid line with a slope of 1.02 is the regression to the cycle~23 data
only.
It is hardly distinguishable from the thick dotted line with a slope of 1.0
expected for a perfect fit.
The corresponding correlation coefficient is 0.94, i.e. the same as found by
\citet{wenzler-et-al-2006a} for the modelled TSI compared to the PMOD
composite for the period since 1992.
We stress that the value of the free parameter, $B_{\rm sat}$, was the
same in both cases.
This means that SATIRE reproduces independent SUSIM data without any further
adjustments, which is yet another success of the model.

In Fig.~\ref{refran}a, we also plot SOLSTICE data \citep{woods-et-al-96}
represented by the blue dashed line.
Note that for comparison sake the SOLSTICE absolute values have been shifted
by $-4.3\times 10^6$\,Wm$^{-3}$nm.
It is clear that at 220--240~nm the model is in a better agreement with the
SUSIM data, even if the correction due to the degradation is not taken into
account, than with the measurements by SOLSTICE, which also show a higher
scatter.

Solar irradiance in the reference range for the period 1996--2002 was also
reconstructed by \citet{krivova-et-al-2006a} using MDI magnetograms and
continuum images (Fig.~2 of that paper).
We have updated their model 
through the beginning of 2006 and combined it with
the KP-based reconstruction shown in Fig.~\ref{refran}.
On the days when both models are available, the preference was given to the
MDI-based values, since they were found to be more accurate
\citep[cf.][]{krivova-et-al-2003a,wenzler-et-al-2004a,wenzler-et-al-2006a}.

Unfortunately, the flat field distortion progressively affecting MDI
continuum images requires a correction of all images recorded after
approximately 2005 before they can be employed for the irradiance
reconstructions.
Such a correction is being attempted, but the outcome is not certain and it
seems advisable to complete the reconstruction instead of waiting an unknown
length of time.

There are also some gaps in the daily reconstructions, when no magnetograms
and continuum images were recorded, in particular, in the 1970s.
On the other hand, climate models often require solar signal input  with a
daily cadence.
Therefore, we have employed the Mg~II core-to-wing ratio
\citep{viereck-et-al-2004} and the solar
F10.7~cm radio flux \citep{tanaka-et-al-73} in order to fill in the gaps and
to extend the data to 2007.
This has been done by using a linear regression between the irradiance in
the reference range (220--240~nm) and the Mg~II index (the linear
correlation coefficient is $R_c=0.98$) and a quadratic relationship between
the irradiance in the reference range and the F10.7 flux ($R_c=0.92$).
Mg~II and F10.7~cm flux data are obtained from the National Geophysical
Data Center (NGDC; http://www.ngdc.noaa.gov/ngdc.html).

\subsection{UV spectral irradiance}
\label{uv}

In order to extrapolate the SATIRE model based on KP NSO and MDI
magnetograms and continuum images to other UV
wavelengths, we made use of the relations between irradiances, $F_\lambda$,
at a given wavelength, $\lambda$, and in the reference interval, $F_{\rm
ref}$ (220--240~nm).
These relationships in the range 115--410~nm were deduced by
\citet{krivova-et-al-2006a} using daily SUSIM data
recorded between 1996 and 2002.
We have repeated this analysis with the data set extended to 2005, but did
not find any significant difference to the earlier derived values and
therefore employed the relationships from the previous work for
consistency.

Using the calculated irradiances at 220--240~nm and empirical
relationships $F_\lambda/F_{\rm ref}$ vs. $F_{\rm ref}$,
solar UV irradiance  at 115--270~nm was reconstructed for the whole period
1974--2007.
Since the long term uncertainty of SUSIM measurements becomes comparable to or
higher than the solar cycle variation at around 250 and 300~nm, respectively
\citep{woods-et-al-96,floyd-et-al-2003b}, above 270~nm SATIRE is found to be
more accurate than the measurements [\citealt{krivova-et-al-2006a}, cf.
\citealt{unruh-et-al-2008}], 
Therefore spectral irradiance values at these wavelengths are calculated
directly from SATIRE.

\section{Results}
\label{results}

\subsection{Ly-$\alpha$ irradiance}
\label{lya}

\begin{figure}
\includegraphics[width=20pc]{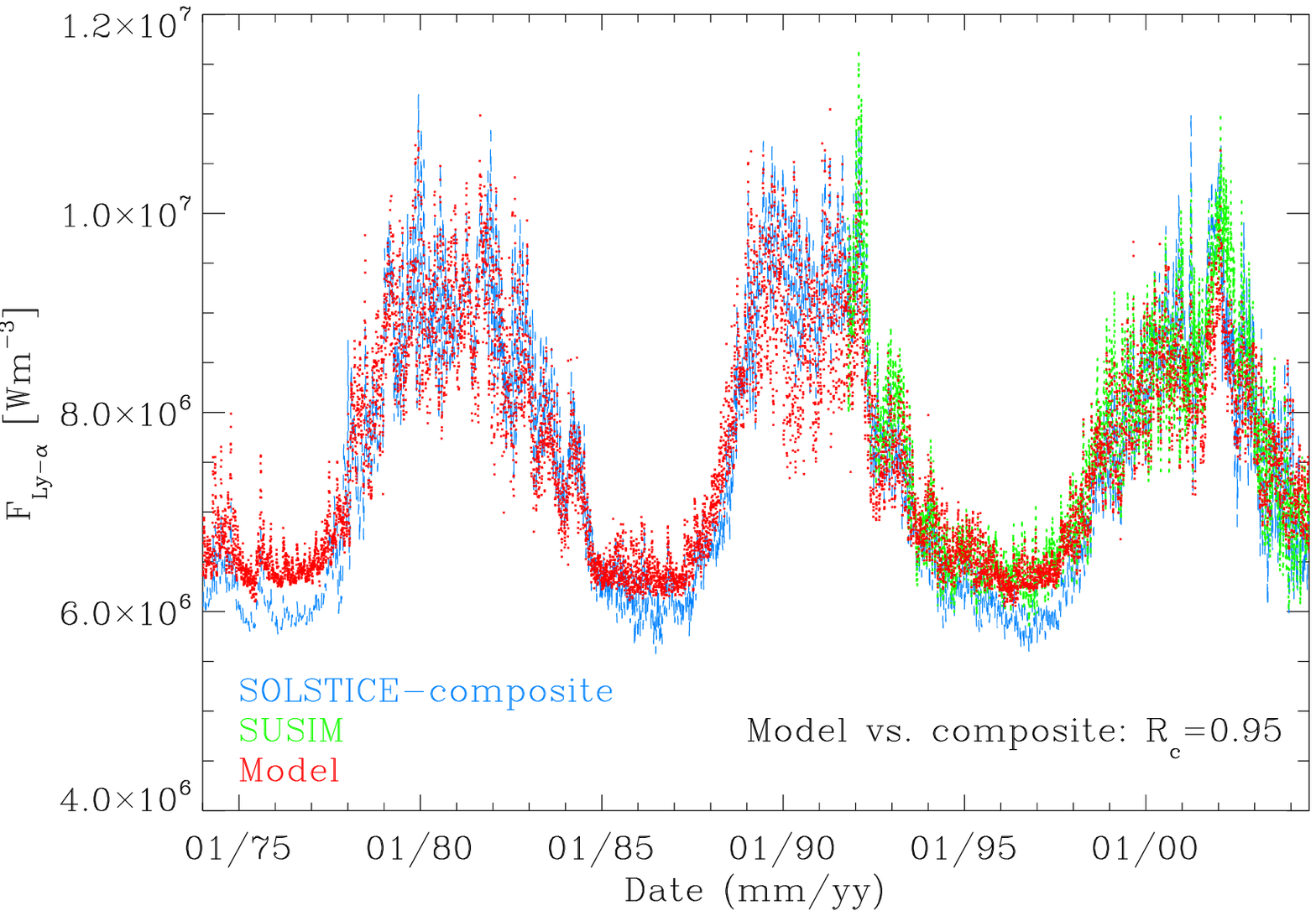}
 \caption[]{\label{lyal}
Solar Ly-$\alpha$ irradiance since 1974: reconstructed by SATIRE (red),
measured by the SUSIM instrument (green) and compiled by
\citet[][blue]{woods-et-al-2000a}.
}
\end{figure}

The Ly-$\alpha$ line is of particular interest not just for its prominence
in the solar spectrum and its importance for the Earth's upper atmosphere,
but also because for this line a composite of measurements is available for
the whole period considered here.
In Fig.~\ref{lyal} we compare the reconstructed solar Ly-$\alpha$ irradiance
(red) with the composite time series (blue) compiled by
\citet{woods-et-al-2000a}.
The latter record comprises the measurements from the Atmospheric Explorer E
(AE-E, 1977--1980), the Solar Mesosphere Explorer (SME, 1981--1989), UARS
SOLSTICE (1991--2001), and the Solar EUV Experiment (SEE) on TIMED
(Thermosphere, Ionosphere, Mesosphere Energetics and Dynamic Mission
launched in 2001).
The gaps are filled in using proxy models based on Mg core-to-wing and F10.7
indices, and the F10.7 model is also used to extrapolate the data set back
in time.
The UARS SOLSTICE data are used as the reference, and other measurements and
the models are adjusted to the SOLSTICE absolute values.
Although this time series is thus only partly based on direct Ly-$\alpha$
observations, it is the nearest we found to an observational time series to
compare our model with.

For comparison, the SUSIM measurements are also plotted (green).
The model agrees well with the SUSIM data, which confirms that our
semi-empirical technique works well.
Note that there is no change in the behavior around  the minimum in 1996.
This is yet another indication of the instrumental origin of the jump in the
absolute values seen in SUSIM's irradiances at 220--240 and many other
wavelengths \citep[see][]{krivova-et-al-2006a}.

As Fig.~\ref{lyal} shows, there is some difference (about 5\%) in the
magnitude of the Ly-$\alpha$ solar cycle variations between SOLSTICE and SUSIM.
Since our model agrees with SUSIM (by construction), a difference of this
magnitude remains also between our model and SOLSTICE.
Other than that, the model agrees with the completely independent composite
time series very well, with a correlation coefficient of 0.95 (remember that
the free parameter of the SATIRE model was fixed from a comparison with the
PMOD composite of the TSI and not varied to fit the UV data).

The solar Ly-$\alpha$ irradiance has also been modelled by
\citet{haberreiter-et-al-2005} using the filling factors derived from the MDI
and KP NSO magnetograms and continuum images in combination with the
brightness spectra for the quiet Sun, sunspots and faculae calculated with
their NLTE code COSI.
The calculated variability was about a factor of 2 lower than the measured
one.
NLTE calculation are, in prinicipal, better suitable for calculations of the
solar UV irradiance and they have recently made significant progress
\citep[e.g.,][]{fontenla-et-al-2006,fontenla-et-al-2007,haberreiter-et-al-2008}.
Their complexity and the number of processes to be accounted for
do not, however, as yet allow an accurate reconstruction of
the solar spectral irradiance over broader spectral
ranges and longer periods of time.

\subsection{Solar UV irradiance at 115--400~nm in 1974--2007}
\label{res_spectrum}

\begin{figure}
\includegraphics[width=20pc]{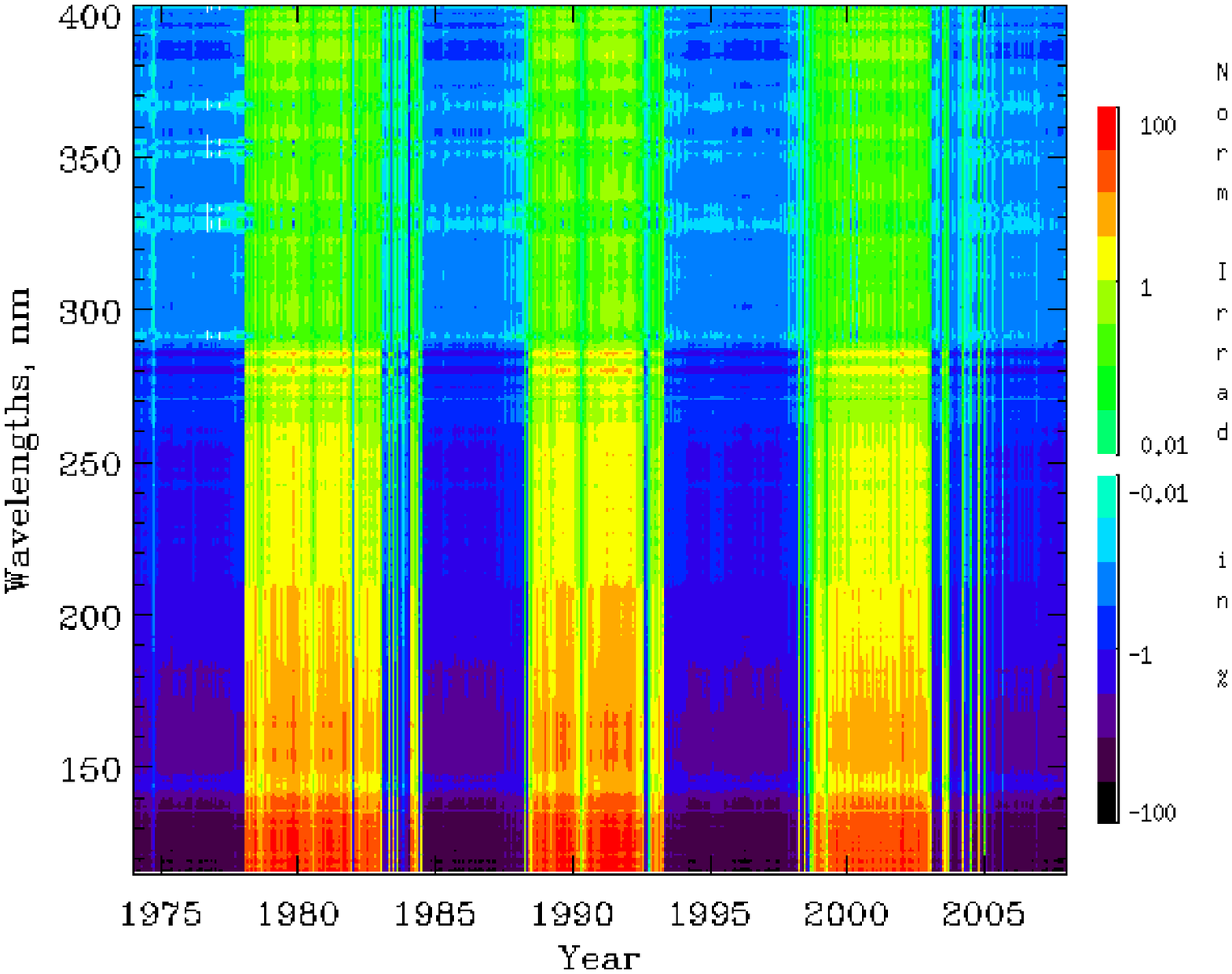}
 \caption[]{\label{fig_uv_sp}
The reconstructed solar UV irradiance at
115--400~nm in the period 1974--2007
normalized to the mean at each wavelength over the whole period of time.
}
\end{figure}

\begin{figure*}
\begin{center}
\includegraphics[width=20pc]{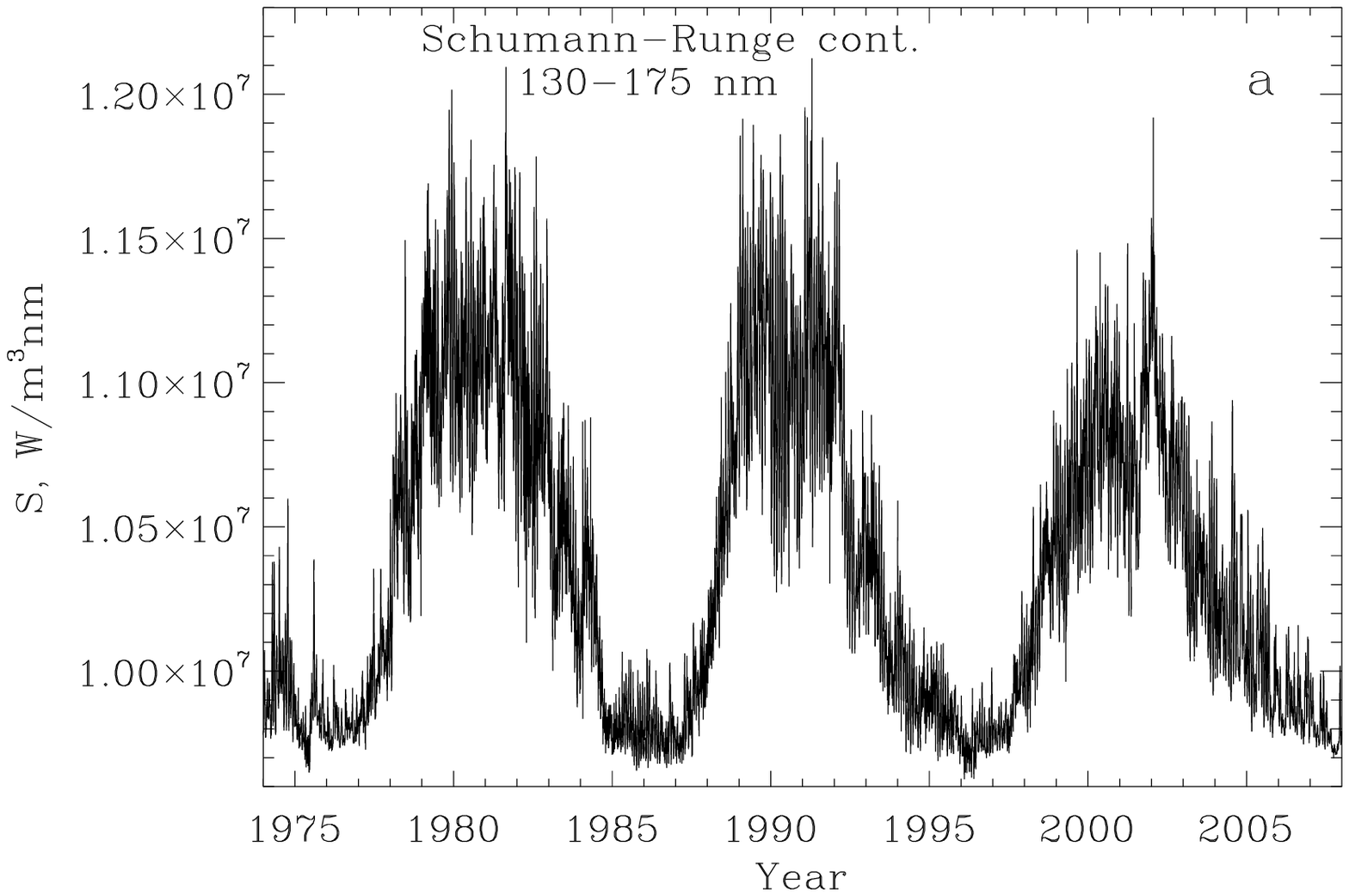}%
\includegraphics[width=20pc]{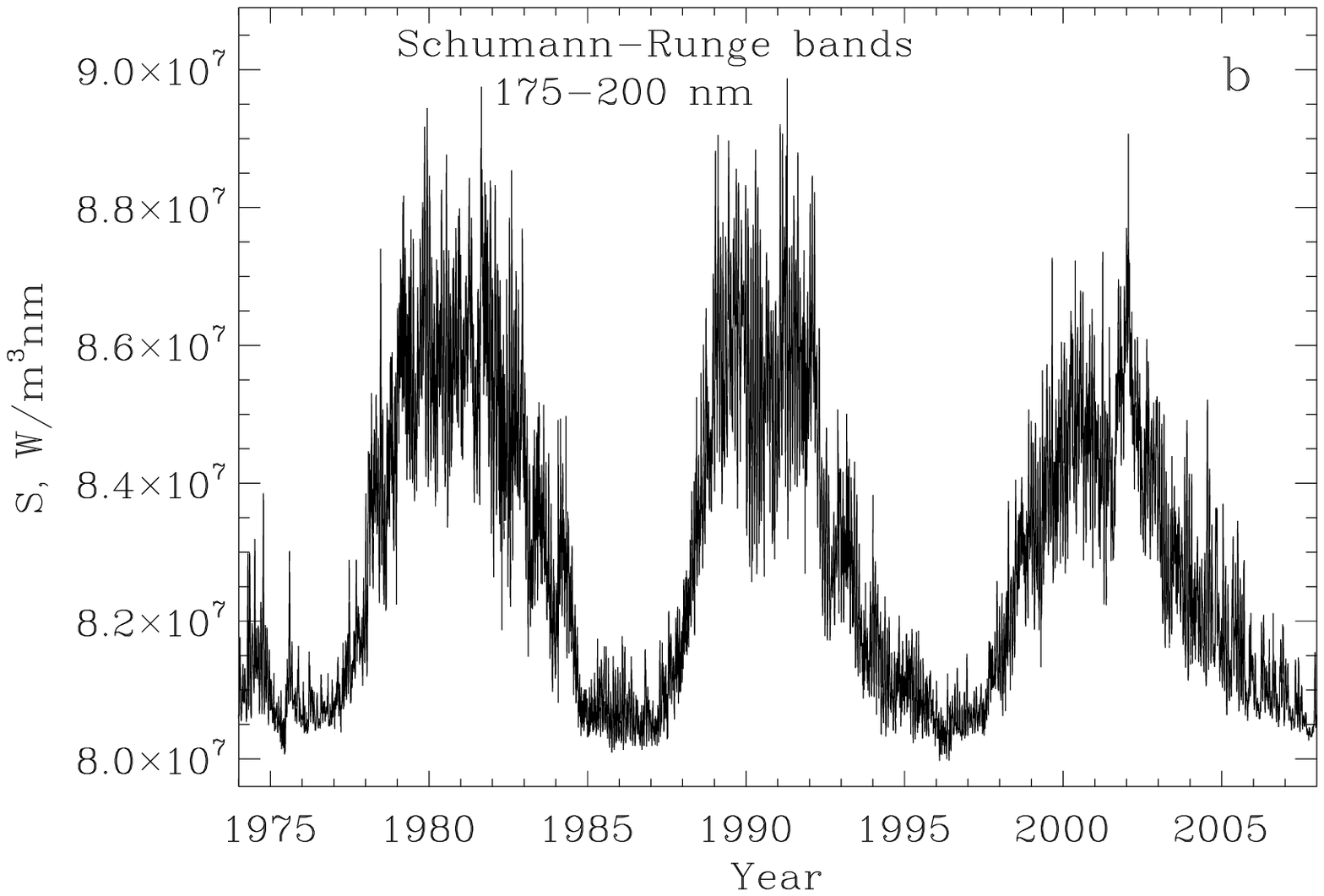}
\includegraphics[width=20pc]{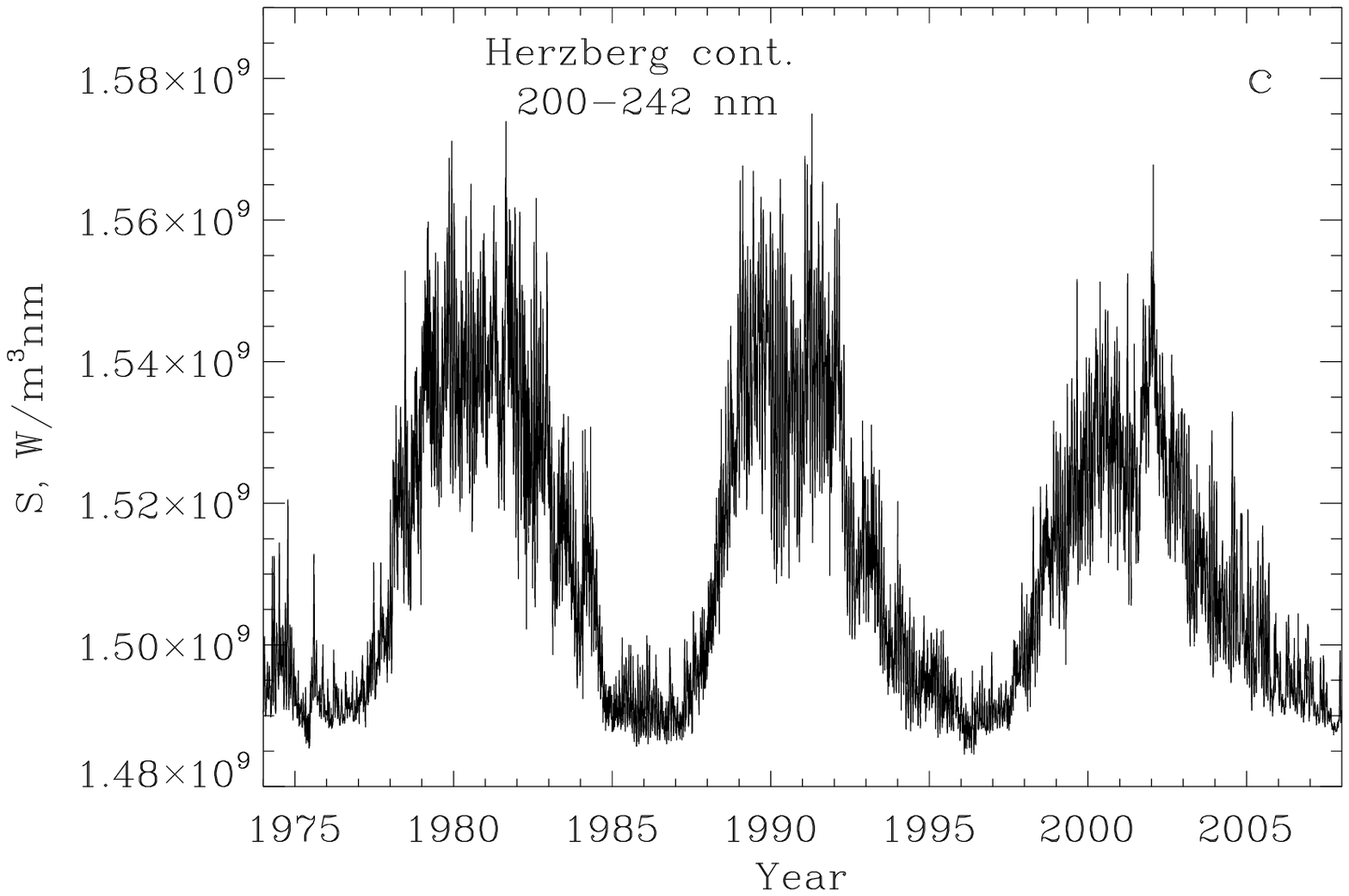}%
\includegraphics[width=20pc]{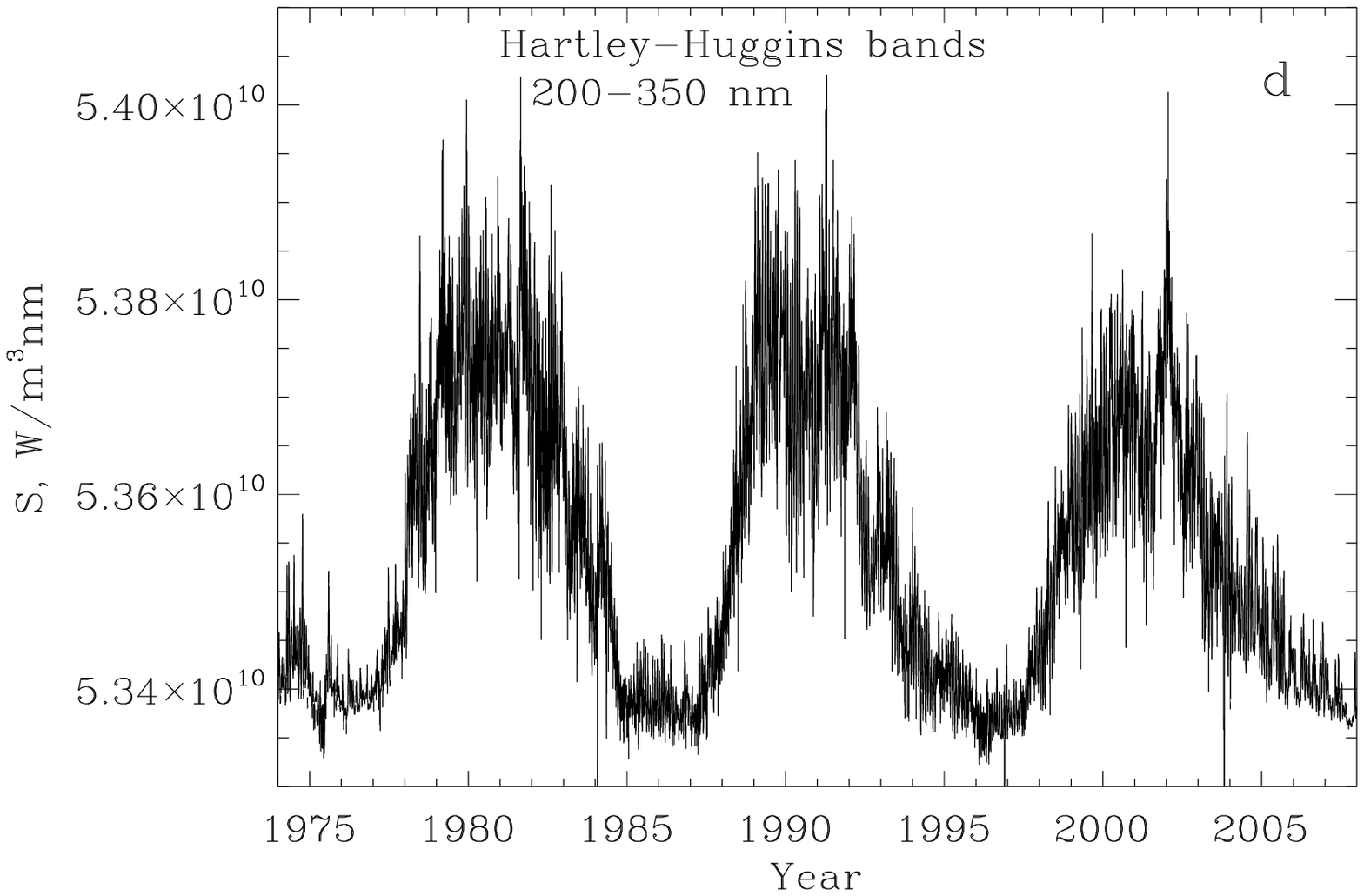}
\end{center}
 \caption[]{\label{fig_uv_bands}
Reconstructed solar irradiance in the period 1974--2007 integrated over the
wavelength ranges: (a) 130--175~nm (Schumann-Runge continuum), (b)
175--200~nm (Schumann-Runge bands), (c) 200--242~nm (Herzberg continuum),
and (d) 200--350~nm (Hartley-Huggins bands).
}
\end{figure*}

Figure~\ref{fig_uv_sp} shows the reconstructed solar UV irradiance in the
range 115--400~nm over the period 1974--2007 (i.e. covering cycles 21--23),
normalized to the mean at each wavelengths over the complete time
period.
At all considered wavelengths, the irradiance changes in phase with the
solar cycle, in agreement with recent results based on 4~years of SIM/SORCE
measurements \citep{harder-et-al-2009}.
The variability becomes significantly stronger towards shorter wavelengths:
from about 1\% over the activity cycle at around 300~nm to
more than 100\% in the vicinity of Ly-$\alpha$.

Figure~\ref{fig_uv_bands} shows the solar UV irradiance integrated over
spectral ranges of particular interest for climate studies as a function of
time: (a) 130--175~nm, (b) 175--200~nm, (c) 200--242~nm, and (d)
200--350~nm.
Solar radiation at 130--175~nm (Schumann-Runge continuum) is completely
absorbed in the thermosphere.
Over activity cycles 21--23, solar radiative flux in this spectral range
varied by about 10--15\% (Fig.~\ref{fig_uv_bands}a), i.e. by more than a
factor of 100 more than solar cycle variations in the solar total energy
flux (total solar irradiance).
In the oxygen Schumann-Runge bands (175--200~nm) and Herzberg continuum
(200--242~nm), important for photochemical ozone production and destruction
in the stratosphere and mesosphere, solar irradiance varied on average by
about 5--8\% (Fig.~\ref{fig_uv_bands}b) and 3\% (Fig.~\ref{fig_uv_bands}c),
respectively.
In the Hartley-Huggins ozone bands between 200 and 350~nm, solar radiation
is the main heat source in the stratosphere.
At these wavelengths, the amplitude of the solar cycle variation is of the
order of 1\%, which is still an order of magnitude stronger than
variations of the total solar irradiance.

The complete data set of the reconstructed solar irradiance at 115--400~nm
over the period 1974--2007 is available as auxiliary material and under
http://www.mps.mpg.de/projects/sun-climate/data.html.

\section{Summary}
\label{summary}

\citet{krivova-et-al-2006a} have developed an empirical technique, which
allows an extrapolation of the magnetogram-based reconstructions of solar
total and spectral irradiance to shorter wavelengths, down to 115~nm.
They applied this technique to obtain variations of solar UV irradiance
between 1996 and 2002.
We have now extended their model to both earlier and more recent times.
Thus we provide a reconstruction of the solar UV irradiance spectrum between
115 and 400~nm over the period 1974--2007.
This extends the available observational record by about 1.5~solar cycles,
i.e. roughly doubles the available record.

As a test of the quality of our model, we have compared the reconstructed
solar Ly-$\alpha$ irradiance with the completely independent composite of
measurements and proxy models by \citet{woods-et-al-2000a}.
There is a small (about 5\%) difference in the solar cycle amplitude
between our model and that composite.
This difference is also present between the SUSIM and SOLSTICE data, which
are the reference sets for the model and the composite, respectively.
Aside from that, the modelled and composite records closely agree with each
other.

Solar UV irradiance varies in phase with the solar cycle at all wavelengths
between 115 and 400~nm, in agreement with the recent finding of
\citet{harder-et-al-2009} based on SIM/SORCE measurements over 2004--2007.
The relative amplitude of the variations grows with decreasing
wavelength. In the wavelength regions important for studies of the
Earth's climate (e.g., Ly-$\alpha$ and oxygen absorption continuum and bands
between 130 and 350~nm), the relative variation is one to two orders
of magnitude stronger than in the visible or if integrated over all
wavelengths (i.e. TSI).

SATIRE-based reconstructed UV irradiance in the spectral range 115--400~nm
between January 1, 1974 and December 31, 2007 is available as auxiliary
material and under http://www.mps.mpg.de/projects/sun-climate/data.html.

%  ACKNOWLEDGMENTS

\begin{acknowledgments}
We thank L.~Floyd for providing SUSIM data and valuable comments and
V.~Holzwarth for helpful discussions.
The composite Lyman~$\alpha$ time series was retrieved from the LASP ftp
server (laspftp.colorado.edu).
This work was supported by the Deutsche Forschungsgemeinschaft, DFG project
number SO~711/2 and by the WCU grant No.~R31-10016 funded by the Korean
Ministry of Education, Science and Technology.
We also thank the International Space Science Institute (Bern) for
giving us the opportunity to discuss this work with the great international
team on ``Interpretation and modelling of SSI measurements''.
\end{acknowledgments}

%% ------------------------------------------------------------------------ %%
%
%  REFERENCE LIST AND TEXT CITATIONS
%
% If you use BiBTeX for your References, please produce your .bbl
% file and copy the contents into your paper here.
%

%\input{jourmnem}
%\bibliographystyle{agu08}  
%\bibliography{sun.bib}
%\input uv74.v3.bblm

\end{article}

% When submitting articles through the GEMS system:
% COMMENT OUT ANY COMMANDS THAT INCLUDE GRAPHICS.

% Figure captions go below this illustration; Table captions go above tables

\end{document}